\documentclass[onesided,10pt]{article}                                                         
\usepackage[left=2.5cm,right=2cm,top=3cm,bottom=2.5cm]{geometry}
\makeatletter
\renewcommand{\fnum@figure}{Fig. \thefigure}
\makeatother

\usepackage{color}
\usepackage{pdfpages}
\usepackage{cite}
\usepackage{algorithm}
\usepackage{algpseudocode}
\usepackage{amsmath}
\usepackage{balance}
\usepackage{multirow}
\usepackage{multicol}   
\usepackage{siunitx}
 
\usepackage{graphicx}
\usepackage[bottom]{footmisc}

\usepackage{epstopdf,float,amsfonts}
\usepackage{epsfig}
\newcommand{\qed}{\hfill \ensuremath{\square}}

 \title{
 \LARGE \bf Structural Impact of Grid-Forming Inverters on Power System Coherency}
 
\author{Sayak Mukherjee$^\dagger$, Ramij Raja Hossain$^\dagger$, Kaustav Chatterjee$^\dagger$, \\Soumya Kundu, Kyung-Bin Kwon, Sameer Nekkalapu, Marcelo Elizondo\\
\textit{Pacific Northwest National Laboratory}\footnote{The Pacific Northwest National Laboratory is operated for the U.S. Department of Energy by Battelle Memorial Institute under Contract DE-AC05-76RL01830. This research is supported by the E-COMP (Energy System Co-Design with Multiple Objectives and Power Electronics) Initiative. $\dagger$Equal contributions, Corresponding Email: sayak.mukherjee@pnnl.gov  }\\
Richland, WA, USA. 
}

\begin{document}
\date{}
\maketitle
\begin{abstract}
\noindent This paper addresses the following fundamental research question: how does the integration of grid-forming inverters (GFMs) replacing conventional synchronous generators (SGs) impact the slow coherent eigen-structure and the low-frequency oscillatory behavior of future power systems? Due to time-scale separated dynamics, generator states inside a coherent area synchronize over a fast time-scale due to stronger coupling, while the areas themselves synchronize over a slower time scale. Our mathematical analysis shows that due to the large-scale integration of GFMs, the weighted Laplacian structure of the frequency dynamics is preserved, however, the entries of the Laplacian may be significantly modified based on the location and penetration levels of the GFMs. This can impact and potentially significantly alter the coherency structure of the system. We have validated our findings with numerical results using the IEEE 68-bus test system.
\end{abstract}


\normalsize

\section{Introduction}
\label{sec:intro}

Increasing growth of renewable energy resource integration in response to global climate challenges has led to an increased presence of inverter-interfaced resources within contemporary power grids. Presently, the forefront of inverter technology mainly consists of two control methodologies: grid-following (GFL) and grid-forming (GFM) as outlined in \cite{pattabiraman2018comparison,lin2020research}. GFLs function as current sources, managing the AC-side current injected into the grid while synchronizing with the grid voltage phase angle through a phase-locked loop (PLL). In contrast, advanced grid-forming (GFM) inverters \cite{gfm_auto, lasseter2019grid} replicates the fundamental behaviors of synchronous generators (SGs) enabling the regulation of system voltages and frequencies. 
GFM technology effectively supports grid stability by autonomously controlling voltage and frequency at the points of interconnection with the transmission system. 
Among different design approaches for GFMs, the droop control stands out as a well-established solution \cite{wei}.
This method involves adhering to the active power-frequency ($P-f$) and reactive power-voltage ($Q-V$) droop curves and guides the power-sharing during steady-states as well as transients. The synchronous generator-esque characteristics have been motivating operators for large deployment of GFM-interfaced renewable resources in future grid \cite{lasseter2019grid}.

From the system theoretic viewpoint, the integration of low-inertia GFM-based resources inherently impacts the dynamics and eigenstructure of the underlying power systems. In small signal stability of power system dynamics, slow coherency is a fundamental characteristic, which results from the time-scale separation of synchronous generator's dynamics \cite{ chowpower, chowbook, podmore}. Generators with strong coupling tend to oscillate in unison, synchronizing rapidly to form slow coherent groups or clusters. Conversely, these groups sway against each other and synchronize over a slower time scale due to weaker coupling. Coherency has been used in operations for transmission planning for dynamic equivalencing, controlled islanding, oscillation damping control, etc \cite{wu,island,ACcontrol}. Few articles have explored the impact of wind generation on coherency \cite{mukherjee2020modeling, khalil2018power}, yet the foundational theory regarding the influence of grid-forming resources on coherency remains unresolved. Keeping the view of futuristic large integration of grid-forming technology \cite{chatterjee_agm}, it is pertinent to fill the gap in theorizing how these resources may impact the coupled-oscillation structures of the grid in the low-frequency ranges (less than $2$ Hz). The implications of this study are multi-fold. The coherent structures in presence of grid forming inverters will help the operator identify the power transfer paths that are influenced by the inter-area modes which many times are poorly damped. Therefore, the operator would be well-guided to design damping control infrastructures with the modified coherent structure information as prescribed by our studies in this work \cite{chatterjee_agm}. Moreover, the dynamic equivalents can also be effectively computed once the structural impact of these inverter dynamics has been captured in the coherent couplings.   

This paper presents the mathematical system description for a power system with droop-based GFMs replacing a fraction of the conventional synchronous generators. 
Thereafter, the paper provides a detailed analytical treatment of how such a heterogeneous resource mix manifests in the small-signal oscillation behavior of the system. We show that the dynamics of the GFMs help to preserve the weighted Lapalcian-like structural behavior for the integrated system; however, the effective coupling between generation resources can lead to considerable perturbations in the slow eigenspace. This, in turn, alters the coherent oscillatory responses, leading to shifting of the cluster boundaries. 
We confirm our propositions with numerical simulations validated on the modified IEEE $68$-bus system with GFMs added at selected locations.

\par The remainder of the paper is structured as follows. Section \ref{sec:modeling} presents the mathematical description of the power system with the models of the synchronous generators and the droop-controlled GFMs. Building on this, in Section \ref{main_results}, we present the analytical results on the impacts of GFMs on the structure of the system Laplacian and coherency. Section \ref{sec:results} presents a case study on an IEEE test system to verify the analytical claims and propositions in Section \ref{main_results}. The concluding remarks are discussed in Section \ref{sec:conclusion}.

\section{System Model: GFM Integration Replacing Synchronous Resources}
\label{sec:modeling}
\begin{figure}[htbp!]
	\centering
	\includegraphics[width=0.5\linewidth]{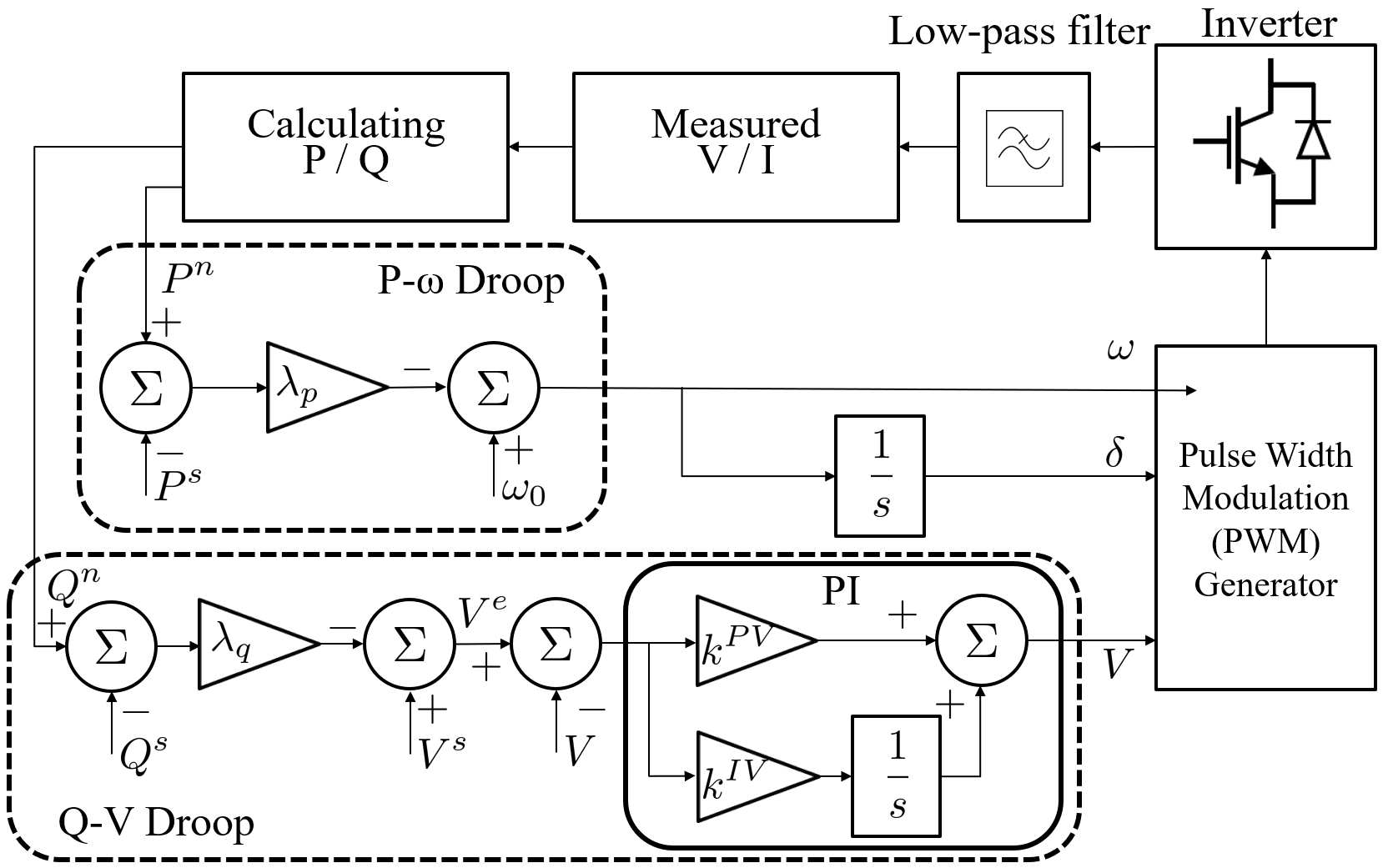}
	\caption{GFM dynamics based on droop controls.}
	\label{fig:dynamics}
    \includegraphics[width=0.5\linewidth]{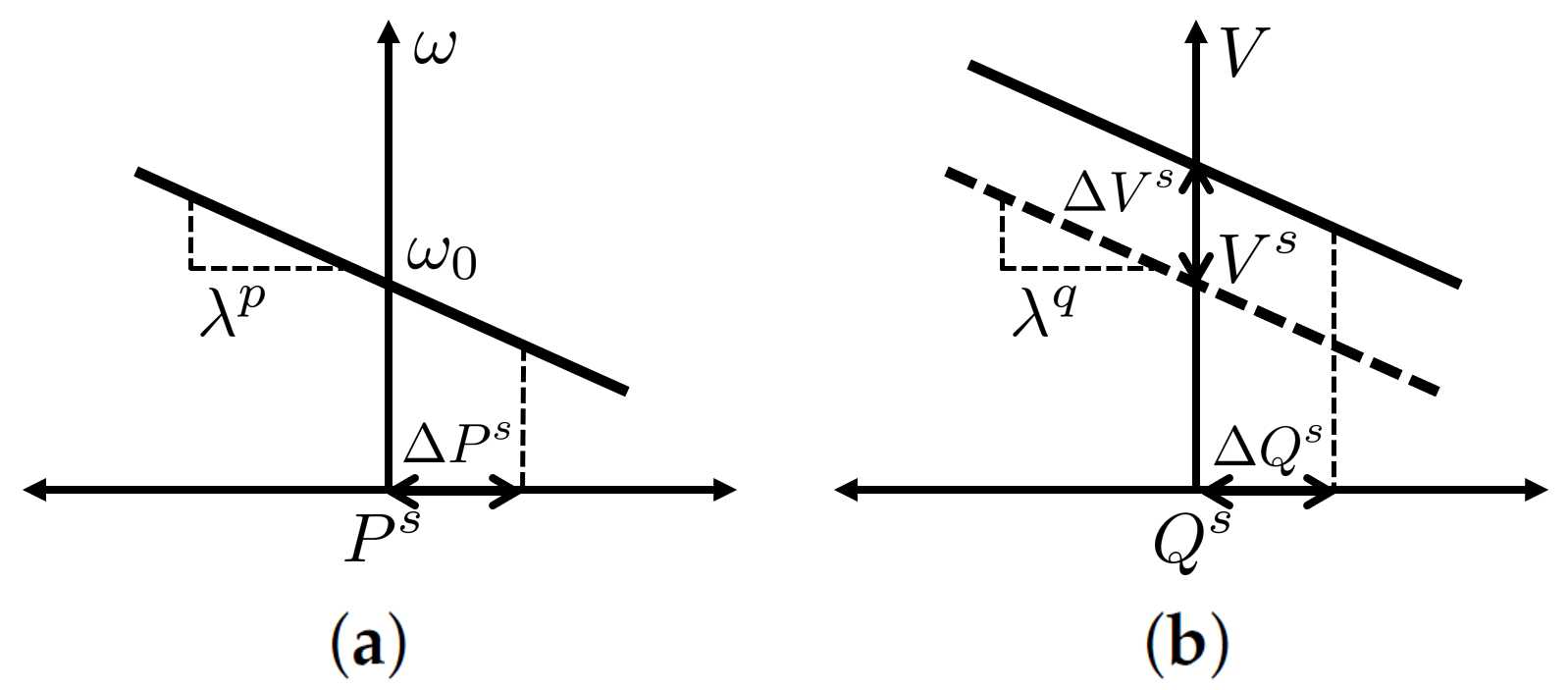}
	\caption{(a) $P-\omega$ and (b) $Q-V$ droop characteristics.}
	\label{fig:droop}
 \vspace{-5pt}
\end{figure}

Consider a power grid composed of $N$ buses. Let $N_r$ be the number of resource buses with either SG or GFM-based resources. The behavior of synchronous generator \( i \) within bus \( i \) is described by the following swing equations \cite{kundur}:
\begin{subequations}
\label{eq:sg}
\begin{align}
&\dot{\delta}_i = \omega_i - \omega_0,\\
&\dot{\omega}_i = \frac{1}{M_i}\big[D_i(\omega_0-\omega_i) + P_i - P_{ei}\big],
\end{align} 
\end{subequations}
\noindent \textcolor{black} Here, \( \delta_i \), \( \omega_i \), \( P_i \), and \( P_{ei} \) denote the angle, frequency, input mechanical, and generated electrical powers of synchronous generator \( i \). Parameters \( \{M_i, D_i\} \) represent the inertia and damping coefficients, respectively. The swing equations represent the dynamics of a SG \cite{kundur} and can be compactly written as, $\dot{x}_s = f_{\mbox{sync}} (x_s, V)$, where $f_{\mbox{sync}}(.)$ denotes the functional form of the dynamics with $V = [V_{1_{Re}}, \dots, V_{N_{Re}}, V_{1_{Im}}, \dots, V_{N_{Im}}]$, and $x_s$ denotes the SG states.  
Initially, we consider all the resources to be SGs; thereafter, we replace $p$ SGs with GFMs having equal ratings. Consequently, after this modification, the number of SGs in the system is reduced to $N_r - p$ indexed in the set $\mathcal{G}$, with $p$ newly added GFMs with indices in the set $\mathcal{F}$. 
For each GFM $j \in \mathcal{F}$, we consider the following dynamics that are also depicted in Fig.~\ref{fig:dynamics} and Fig.~\ref{fig:droop} \cite{wei, soumya_droop, kwon_gfm} : 
\begin{subequations}
\label{eq:gfm}
\begin{align}
&\dot{\delta}_j = \omega_j - \omega_0,\\ &\dot{\omega}_j = \frac{1}{\tau_j}\big[\omega_0 - \omega_j +\lambda^p_j (P^{s}_j - P^n_j)\big],\\ &\dot{V}^e_j = \frac{1}{\tau_j}\big[V^{s}_j - V^e_j - V_j + \lambda^q_j (Q^{s}_j - Q^n_j)\big],\\ &\dot{E}_j = k^{pv}_j \dot{V}^e_j + k^{iv}_j V^e_j,
\end{align} 
\end{subequations}
where $\lambda_j^p, \lambda_j^q, \tau_j$ are the active and reactive power droops and droop time constant, $k^{pv}_j$ and $k^{iv}_j$ are respectively the proportional and integral gains in the Q-V droop loop, $\tau_j$ is the time-constant associated with measurement filter, $V^s$, $P^s$, and $Q^s$ are the set-points for voltage, active power and reactive power, respectively. The state vector for each GFM $j$ becomes $x_j := [\delta_j, \omega_j, V^e_j, E_j]$. We stack these states together and denote the state-vector as $x_f:= [\{ x_j\}_j]$. Their dynamics can then be written as $\dot{x}_f = f_{\mbox{gfm}} (x_f,  V).$ 

Next, the active and reactive power flow balance at any bus $j,\; j=1,\dots,m$ can be written as,
\begin{subequations}
\label{eqn:load_flows}
\begin{align}
\label{eqn:load_active} 0=&{P_{ej}} - {\rm{Re}}\left\{
{\sum\limits_{k = 1,k \ne j}^N { {{V_{j}}}{{\left(
{V_{jk}}{B_{jk}} \right)}^*}} } \right\} - V_j^2{G_j},\\
\label{eqn:load_reactive4} 0=&{Q_{ej}} - {\mathop{\rm
Im}\nolimits} \left\{ {\sum\limits_{k = 1,k \ne j}^N
V_{j}{{{\left( {V_{jk}}{B_{jk}}\right)}^*}} }
\right\} - V_j^2{B_j},
\end{align}
\end{subequations}
where $G_j$ and $B_j$ represent the conductance and susceptance, respectively, of the shunt load at bus $j$ when considering line charging. In the case of lossless transmission lines, $B_{jk}$ indicates the susceptance of the tie-line linking bus $j$ to bus $k$. The powerflow equations can be expressed with the functional form captured by $g(\cdot): 0 = g(x_s, x_f, V)$.

Next, we will investigate the integrated small-signal frequency dynamics, aiding in the characterization of perturbations within the Laplacian dynamics, ultimately influencing the slow coherency. 

\section{Main Results: Impacts of GFMs on System Laplacian and Slow Coherency}
\label{main_results}

\subsection{Structure-preserving Laplacian Sub-Matrix} 
To extract the Laplacian sub-matrix of the state dynamics, we examine the frequency dynamics influenced by the angles associated with the resources. Consider a vector that concatenates all angles for SGs denoted as $\delta_s$ and for GFMs as $\delta_f$. Similarly, vectors for frequencies are denoted as $\omega_s$ for SGs and $\omega_f$ for GFMs. We can then express the frequency dynamics concisely for SGs as follows:
$
\label{eqn:freq_gen}
    M\dot{w}_s = \bar{f}_{\mbox{sync}} (\delta_s, V),
$
where $\bar{f}_{\text{sync}}(.)$ represents the part of the complete functional form corresponding to the frequency dynamics. The SG angles influence the dynamics, as the electrical active power from SG $i$ is defined by $${P_{si}} =\frac{{{E_i}}}{{x_{di}'}}\left( {{V_{{i_{Re}}}}\sin {\delta _i} - {V_{{i_{Im}}}}\cos {\delta _i}} \right).$$ Now, for the GFMs, we can similarly express this in a concise form as,
$
\label{eqn:freq_gfm}
    M_f\dot{w}_f = \bar{f}_{\mbox{gfm}} (\delta_f, E_f, V),
$
where $\bar{f}_{\text{gfm}}(.)$ represents a segment of the overall dynamics related to the frequency evolution, and $E_f$ denotes the stacked internal voltage states for inverters. The equivalent inertia, $M_f$, offered by the grid-forming inverter unit depends on the GFM droop control parameters: $M_f = \text{blkdiag}(\{\frac{\tau_j}{\lambda^p_j}\}_j)$. Also, the powerflow equations in the functional form can be slightly expanded to include the additional state variables, such that
$
\label{eqn:pf_compact2}
    0 = g(\delta_s, \delta_f, E_f, V).
$
At a stable operating point, linearizing the differential equations from SG and GFM dynamics, and the algebraic equations from powerflow, we have \vspace{-0.1cm}
\begin{align} 
    & M \Delta \dot{w}_s = A_{11} \Delta \delta_s + A_{12} \Delta V, \label{eq1}\\
     & M_f \Delta \dot{w}_f = A_{21} \Delta \delta_f + A_{22} \Delta V, + A_{23} \Delta E_f, \label{eq2}\\
      & 0 = A_{31} \Delta \delta_s + A_{32} \Delta \delta_f + A_{33} \Delta V + A_{34} \Delta E_f, 
\vspace{-0.1cm} 
\end{align} 
where the matrices $A_{ij}$ denote the Jacobians corresponding to the argument variables. 
Further, we define, 
\noindent $A_1 = blkdiag(A_{11}, A_{21}), A_2 = [A_{12}; A_{22}], M_e = blkdiag(M, M_f), A_3 = [A_{31} A_{32}], A_4 = [\textbf{0}; A_{34}]$. 
Following which we may write, \vspace{-0.2cm}
\begin{align}
    \Delta V = -A_{33}^{-1} A_3 \begin{bmatrix}
        \delta_s \\ \delta_f
    \end{bmatrix}  -A_{33}^{-1} A_{34} \Delta E_f.
\end{align}
Using this expression in \eqref{eq1}, and \eqref{eq2}, we will have,
\small
\begin{align}
    M_e \begin{bmatrix}
        \Delta \dot{w}_s \\ \Delta \dot{w}_f
    \end{bmatrix} = (A_1 - A_2A_{33}^{-1} A_3 )\begin{bmatrix}
        \delta_s \\ \delta_f
    \end{bmatrix} + (A_4 - A_2A_{33}^{-1} A_{34} )\Delta E_f.
\end{align}
\normalsize
Therefore, the Laplacian matrix, $L$, for this Kron-reduced coupled SG-GFM interaction network is captured by
\begin{align}
    L = (A_1 - A_2A_{33}^{-1} A_3 ), 
    \vspace{-0.2cm}
\end{align}
and the sub-matrix weighted by inertia that captures the impact of angles on the frequency dynamics is expressed as $\bar{L} = M_e^{-1}L$. The eigenstructure of this weighted Laplacian matrix compared to that of the all-synchronous generator case determines the extent to which the addition of GFMs alters the dynamic coupling between the generating resources. \vspace{0.1cm}
\par

\vspace{.4 cm}
\noindent \textbf{Lemma 1.} \textit{GFMs will preserve the Laplacian structure of the integrated sub-matrix $L$, ensuring that $L \cdot \textbf{1} = \textbf{0}$.}
\par
\noindent \textit{Proof:} The frequency dynamics of GFMs (GFMs) induce behavior resembling synchronous behavior, incorporating modifications to the inertia parameters. The equivalent inertia is determined by the parameter $\frac{\tau_j}{\lambda^p_j}$ for inverter $j$. Consequently, after the admittance matrices are Kron-reduced to capture the all-to-all coupling among generation resources, including SGs (SGs) and GFMs, the active power generated by the $i^{th}$ GFM unit towards another GFM or SG (let's denote it as the $j^{th}$ unit) is equivalently given by $E_iE_jB_{ij}\sin(\delta_i - \delta_j)$, where $B_{ij}$ represents the equivalent Kron-reduced admittance between SG/GFM resources. Therefore, the components of the matrix $L$ will result in: 
\begin{align}
    L(i,j) &= E_iE_jB_{ij}cos(\delta_{i0} - \delta_{j0}) \;\; \mbox{if} \;\;  i \sim j,\\
    &= 0, \mbox{if disconnected.}
\end{align}
Because of the active power-balance structure between the SG/GFMs to supply the loads, we will have,
\begin{align}
    L(i,i) = -\sum_{j \in \mathcal{N}_i} L(i,j).
\end{align}
Therefore, we will have $\textbf{1}$ as one of its eigenvectors with zero eigenvalue resulting in the condition: $L. \textbf{1} = \textbf{0}$. 
\qed
\subsection{Perturbation in Coherent Boundaries} \vspace{-0.05cm}
 Despite replacing the active power generation from SGs (SGs) with GFM-based resources, operational constraints may limit the operator's ability to connect GFM resources only to nearby buses. As a result, the integration of GFMs in place of SGs perturbs the operating points. However, it can be assumed, without loss of generality, that the error introduced to power flow Jacobians due to slight shifts in operating conditions is sufficiently small for practical purposes. If the operator can substitute SG generation with an equal capacity GFM, this would also justify the small perturbation. When considering only conventional synchronous machines integrated into the grid, the linearized model can be expressed as follows:
\begin{align}
    M \Delta \dot{w}_s = A_{11} \Delta \delta_s + A_{12} \Delta V,
      0 = A_{31} \Delta \delta_s + A_{33} \Delta V,
\end{align}
and the Kron-reduced frequency dynamics would lead to,
\begin{align}
    M \Delta \dot{\omega}_s = (A_{11} - A_{12}A_{33}^{-1}A_{31})\Delta \delta_s.
\end{align}
Hence, the inertia-weighted Laplacian for the grid with only SGs is defined as
\begin{align}
    \bar{L}_0 = M^{-1}(A_{11} - A_{12}A_{33}^{-1}A_{31}).
\end{align}
As a result, the perturbation induced by significant GFM integration, resulting in $M_e^{-1}L$, would capture the inherent clustering behavior. Consequently, we present the following proposition regarding the modified coherency behavior.

\vspace{0.4 cm}
\noindent \textbf{Proposition 1.} \textit{Considerable GFM-based replacement can lead to shifts in coherent area partitions through perturbations in the slow eigen-space.}\\
\textit{Proof:} We consider $r$ coherent areas, taking into account the slow coherent behaviors as discussed in \cite{chowbook}. We examine matrices $W_r$ and $\bar{W}_r$, whose columns are the eigenvectors corresponding to the zero eigenvalue and the $(r-1)$ slow eigenvalues of the weighted Laplacians $M_e^{-1}L$ and $M^{-1}\bar{L}$, respectively. In \cite{chowpower}, a Gaussian elimination-based procedure is used to construct permuted versions of $W_r$ and $\bar{W}_r$ by identifying the group reference machines denoted as $\mathcal{W}_r$ and $\bar{\mathcal{W}}_r$. The canonical angles between these two sub-spaces are defined as $\theta_i = \cos^{-1}\sigma_i, i=1,\dots,r$, where $\sigma_i, i=1,\dots, r$ are the $r$ smallest singular values of $\bar{\mathcal{W}}_r^T\mathcal{W}_r$. Let $e_i$ and $\bar{e}_i$ be the $i^{th}$ eigenvalues of $M_e^{-1}L$ and $M^{-1}\bar{L}$, respectively. If there is a gap $|e_r - \bar{e}_{r+1}| > \beta,  \, \beta >0$, then the perturbation in the sub-spaces can be captured by

\begin{align}
||\sin(\Theta)||_F \leq \frac{1}{\beta} ||M^{-1}\bar{L}\mathcal{W}_r - \mathcal{W}_r\Sigma_r||_F,
\end{align} 

\noindent where $\Theta = \text{diag}(\theta_1,\dots, \theta_r)$, and $\Sigma_r = \text{diag}(e_{1},\dots, e_{r} )$ \cite{davis1970rotation,hunter2010performance}. Due to this perturbation, the row vectors of the permuted GFM-integrated slow-subspaces ($\alpha_i$'s) would be perturbed from the all-synchronous case ($\bar{\alpha_i}$'s) as follows \cite{hunter2010performance}:
\begin{align}
    ||\alpha_i - \bar{\alpha}_iQ||_F \leq \frac{1+\sqrt{2}}{\beta}||M_e^{-1}L-M^{-1}\bar{L}||_F,
\end{align}
Here, $Q$ is an orthogonal matrix that minimizes $||\mathcal{W}_r - \bar{\mathcal{W}}_rQ||_F$. However, as discussed in \cite{chowbook}, these row vectors are constrained by the hyperplane 

\begin{align}
\sum_{j=1:r} \alpha_{ij} = 1, \,\,\,i=1,\dots,n.
\end{align}

\noindent Subsequently, following Lemma 1, they will undergo a rotational shift, leading to the formation of a new clustering structure with a substantial change in $||M_e^{-1}L-M^{-1}\bar{L}||_F$. This captures the impact of GFM integration, introducing a perturbation in the slow eigenspace that propagates through the row vectors of the slow subspace, ultimately altering the clustering. \qed




\noindent \textbf{Remark 1} (\textit{time-scale separation}). As the connections among resources within the same coherent areas being more rigid compared to those between different areas, we can still decompose, $L = L^I + \epsilon L^E$, where $L^I$, and $L^E$ are the internal and external coupling matrices, and $\epsilon$ captures the worst-case ratio of tie-line reactances, considering both internal and external connections within the coherent areas. With GFM integrated dynamics the slow variable would be given as $$\frac{\sum_{i \in \mathcal{G}} M_i \Delta \delta_{s_i} + \sum_{i \in \mathcal{F}} M_{f_i} \Delta \delta_{f_i}}{\sum_{i \in \mathcal{G}} M_i + \sum_{i \in \mathcal{F}} M_{f_i}}$$ for a particular modified coherent cluster, and the angle differences from a reference would be the fast variable.


\section{Case Studies and Numerical Validations}
\label{sec:results}
\begin{figure}[h]
\centering
\includegraphics[width=0.6\linewidth]{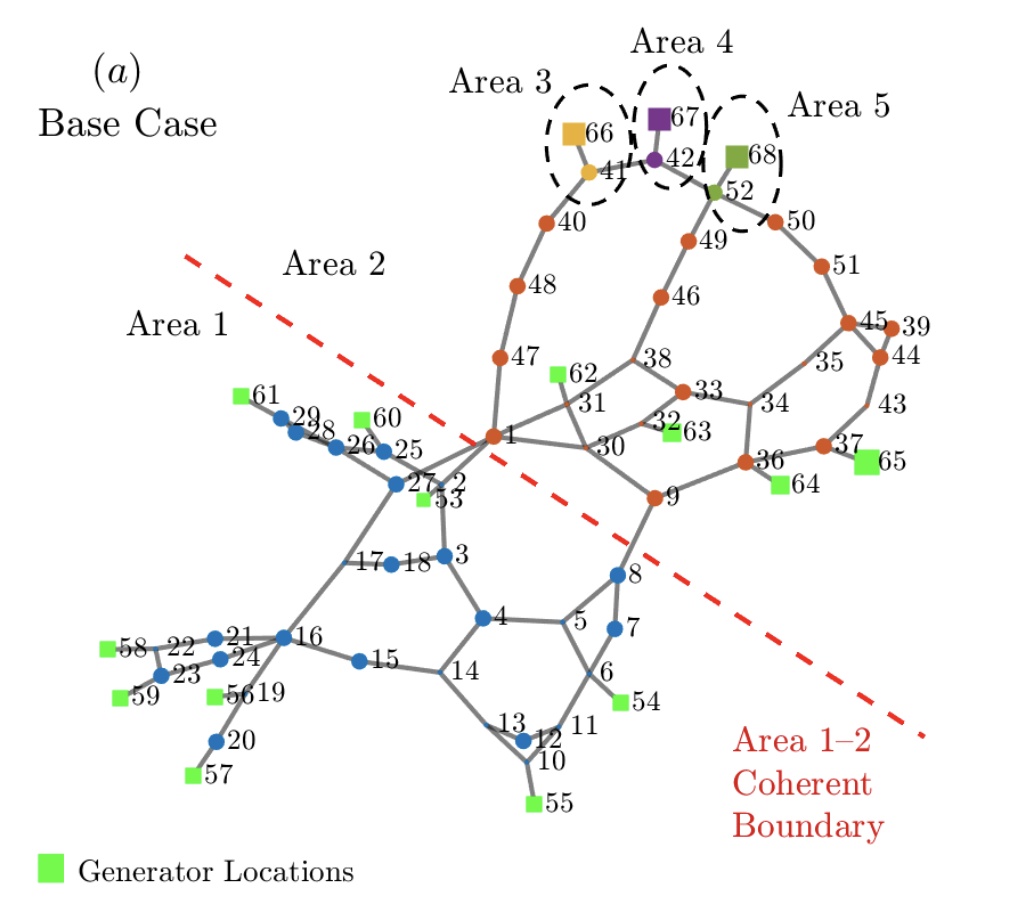}
\caption{Coherent inter-area boundary between areas 1 and 2 for the base case.}
\label{fig:16mc}
\vspace{-0.3 cm}
\end{figure}

This section presents results from numerical studies on an IEEE test system to validate the claims and propositions made in the paper. The positive-sequence fundamental-frequency phasor model of the IEEE 68-bus New York -- New England test system is considered. The base case of the IEEE 68-bus test system has 16 SGs in 5 coherent areas as shown in Fig. \ref{fig:16mc}. The generators are connected to buses numbered between 53 and 68. We focus on the generators in Areas 1 and 2 (see the base case in Table \ref{table:coherent_areas}), with their inter-area or coherent group boundary shown with \textit{red dotted line} in Fig. \ref{fig:16mc}. 

\begin{figure}[t]
\centering 
\includegraphics[width=\linewidth]{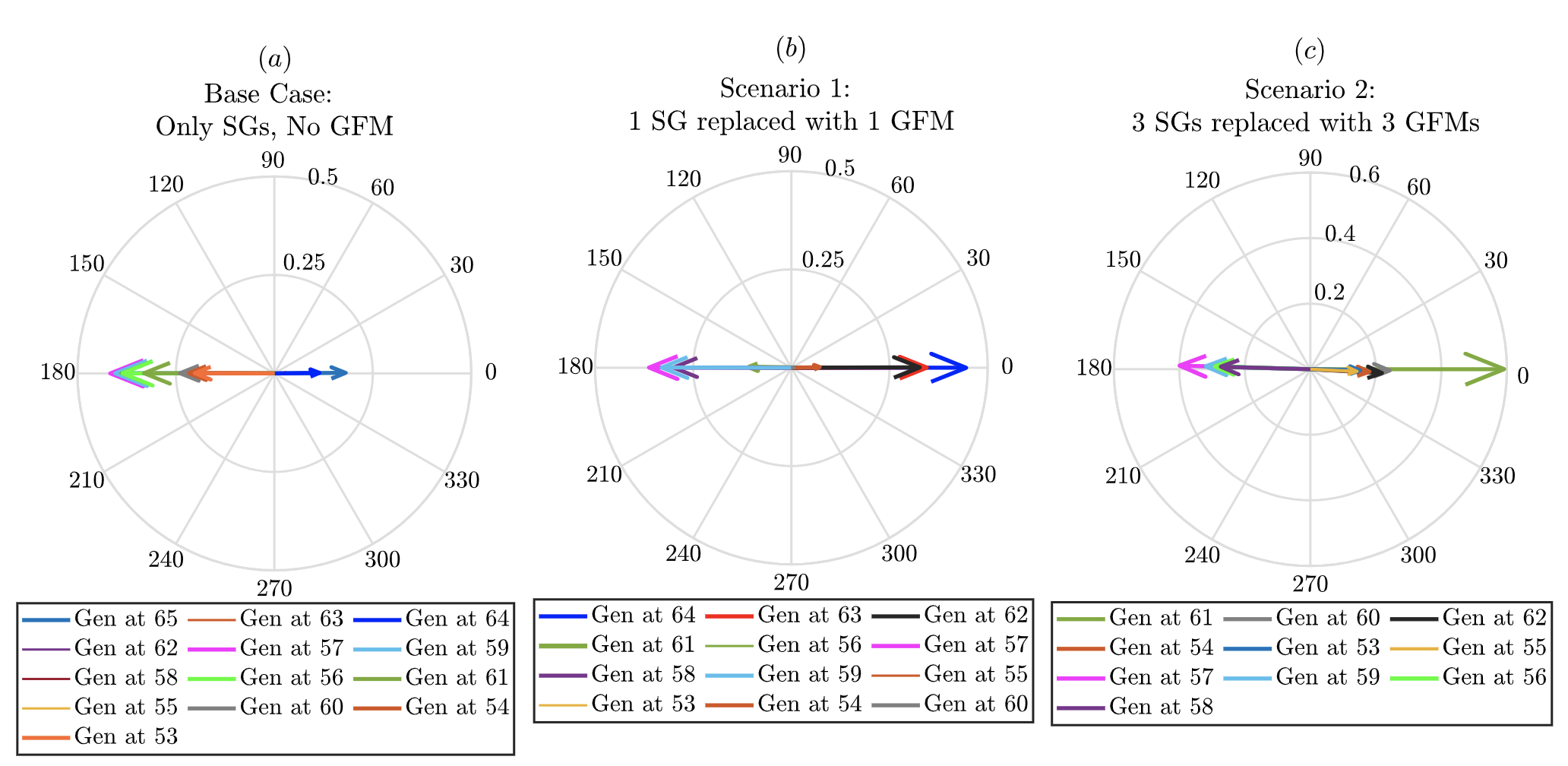}
\caption{Generator modeshapes of the dominant inter-area mode for different test scenarios: $(a)$ Base case, $(b)$ Scenario 1, and $(c)$ Scenario 2.}
\label{fig:ms1}
\end{figure}

\renewcommand{\arraystretch}{1.3}
\begin{table}[b!]
\tabcolsep=0.05 cm
  \centering
  \caption{Summary of the Test Case Scenarios and their Coherent Groups}
  \label{table:coherent_areas}
  \vspace{0.2cm}
  \begin{tabular}{c||c||c||c} 
    \hline
    \textbf{Scenarios} & \textbf{Generators Replaced} & \textbf{Generators in Area 1} & \textbf{Generators in Area 2}\\
    \hline
    (a) Base Case & $-$ & Bus: 53, 54, 55, 56, 57, 58, 59, 60,  61 & Bus: 62, 63, 64, 65 \\ \hline 
    (b) Scenario 1 & Bus: 65 & Bus: 56, 57, 58, 59, 61 & Bus: \textbf{53}, \textbf{54}, \textbf{55}, \textbf{60}, 62, 63, 64 \\ \hline 
    (c) Scenario 2 & Bus: 63, 64, 65 & Bus: 56, 57, 58, 59   & Bus: \textbf{53}, \textbf{54}, \textbf{55}, \textbf{60}, \textbf{61}, 62 \\ \hline
  \end{tabular}
  \vspace{-5pt}
\end{table}

\begin{figure}[t]
\centering
\vspace{-0.4cm}
\includegraphics[width=\linewidth]{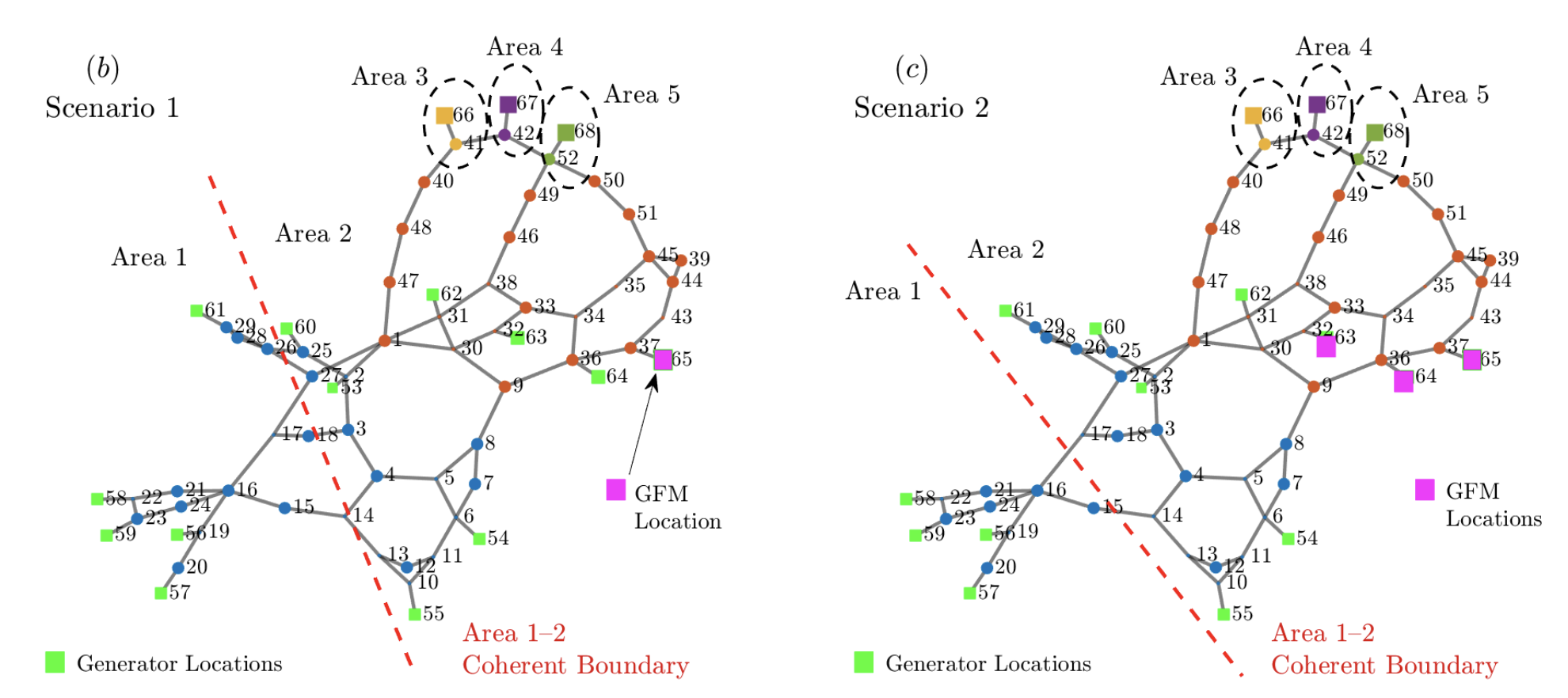}
\caption{Coherent inter-area boundary between Areas 1 and 2 for Scenarios 1 and 2.}
\label{fig:coherent_area_1_2}
\vspace{-0.34cm}
\end{figure}


In this study, we investigate the impact of replacing SGs with GFMs on the coherent groups and their inter-area boundary. To this end, we study the following two scenarios.

\begin{enumerate}
    \item \emph{Scenario 1}: The generator at bus 65 is retired and replaced with a GFM at the adjoining bus 37.
    \item \emph{Scenario 2}: The generators at buses 63, 64, and 65 are retired and replaced with GFMs at their adjoining buses 33, 36, and 37 respectively.
\end{enumerate}

The steady-state power injection set-points for the GFMs are the same as those of the retiring SGs. The total real power injection from the GFMs in each scenario, both in absolute values and as a percentage of the system load (18408 MW), is shown in Table. \ref{table:gfm_inj}. The system model for each case is linearized about the steady-state equilibrium and the inter-area modes obtained from eigenanalysis are shown in Table. \ref{table:interarea_modes}.
\renewcommand{\arraystretch}{1.3}
\begin{table}[h]
    \centering
    \caption{Real Power Injections from the GFMs in Different Scenarios}
    \label{table:gfm_inj}
    \vspace{0.2cm}
    \begin{tabular}{c||c||c} \hline
        \bf{Scenarios} & \bf{Total Real Power from GFMs} & \bf{\% of Total Load} \\ \hline \hline
          1 & 3591 MW &  20\% \\ \hline
          2 & 5941 MW &  33\% \\ \hline
    \end{tabular}
\end{table}

\begin{table}[h]
    \centering
     \caption{Inter-Area Modes for Different Scenarios}
    \label{table:interarea_modes}
    \vspace{0.2cm}
    \begin{tabular}{c|| c | c | c} \hline
        & \bf{Base Case} & \bf{Scenario 1} & \bf{Scenario 2} \\ \hline\hline
                                & 0.41 Hz & 0.47 Hz & 0.49 Hz \\ 
        \bf{Inter-area Modes }  & 0.57 Hz & 0.56 Hz & 0.56 Hz  \\
        {(Modal Frequencies)}  & \bf{0.75 Hz} & \bf{1.14 Hz} & \bf{1.36 Hz}  \\
                                & 0.85 Hz & 0.85 Hz & 0.85 Hz \\ \hline
    \end{tabular}
    \vspace{-5pt}
\end{table}

From Table \ref{table:interarea_modes}, observe that the inter-area modal frequencies 0.4 Hz, 0.57 Hz, and 0.85 Hz remained unchanged from the system modifications. However, the 0.75 Hz mode in the base case changed significantly in Scenarios 1 and 2 from the SG retirement and GFM addition. The relative participation of the retired generators in the inter-area modes can explain this selectivity in modal behavior. The retirement of a large ($\approx$ 20\% of total capacity) generator in Scenario 1 reduces the system inertia, which increases the modal frequency\footnote{For the relationship between single machine-equivalent inertia and modal frequency, please see \cite[Eqn. (12.81)]{kundur}.}.

The system changes in Scenarios 1 and 2 also cause a shift in the inter-area or coherent area boundaries in the system. This is verified by the mode shapes obtained from eigenvector analysis. 
In Fig.~\ref{fig:ms1}, we plot the modeshapes of the generator speeds corresponding to the GFM-sensitive 0.75 Hz inter-area mode for the base case and scenarios 1 and 2 respectively.  Observe that, transitioning from the base case to Scenario 1, the SGs connected to buses 53, 54, 55, and 60 flipped from Area 1 (coherent with SGs at buses 56 $-$ 61) to Area 2 (coherent with SGs at buses 62 $-$ 64). Further, with more SGs replaced in Scenario 2, the change in the slow-coherent oscillation pattern is more pronounced, as shown in Fig.~\ref{fig:ms1}(c). In this case, in addition to the SGs at buses 53, 54, 55, and 60 from Scenario 1, the SG at bus 61 also flipped from Area 1 to 2. The summary of the cases and the generators in the coherent groups are shown in Table \ref{table:coherent_areas}. Figure ~\ref{fig:coherent_area_1_2} shows the transition of coherent area boundary (\textit{red dotted line)} from the base case to Scenario 1, and finally to Scenario 2.

{
To validate our findings on the coherency groups, we compared our results with those obtained from \textit{Slow-Coherency Grouping Algorithm} \cite[Chapter~3.4]{chowpower}. 
The algorithm outcomes for Scenario 1 and 2 are listed in Table~\ref{table:coherent_areas_valid}. Now, comparing Table~\ref{table:coherent_areas} and Table~\ref{table:coherent_areas_valid}, we found that the coherent grouping obtained by the algorithm identically matches with our observation for Scenario 2, where 3 SGs (SGs at buses 63, 64, and 65) are replaced, while there is slight disagreement in Scenario 1, where only SG at bus 65 is replaced. 
We further observed that the mismatches are related to SGs at buses 53, 54, and 60, which have relatively less contribution in interarea mode (see Fig.~\ref{fig:ms1}(b)). Additionally, from Fig.~\ref{fig:coherent_area_1_2} (b) it is evident that these 3 SGs are located close to the interarea boundary; therefore intuitively the grouping algorithm appears less sensitive for those low contributing SGs, causing the mismatch.

\renewcommand{\arraystretch}{1.3}
\begin{table}[h]
  \centering
  \caption{Coherent Groups with \textit{Slow-Coherency Grouping Algorithm} \cite{chowpower}}
  \label{table:coherent_areas_valid}
  \vspace{0.2cm}
  \begin{tabular}{c||c||c} 
    \hline
    \textbf{Scenarios} &   \textbf{Generators in Area 1} & \textbf{Generators in Area 2}\\
    \hline 
  Scenario 1  & Bus: 53, 55, 56, 57, 58, & Bus: \textbf{54},  62, 63, 64 \\ {} & 59, 60, 61 & {} \\\hline 
 Scenario 2  & Bus: 56, 57, 58, 59   & Bus: \textbf{53}, \textbf{54}, \textbf{55}, \textbf{60}, \textbf{61}, 62 \\ \hline
  \end{tabular}
\end{table}

Finally, we provided numerical validation of \textbf{Lemma 1}. Please note that $L \cdot \textbf{1}$ is a vector. Table~\ref{table:L1} shows that the statistical dispersion of the elements of the $L \cdot \textbf{1}$ vector expressed in terms of their (a) mean and standard deviations (Std.), and (b) maximum value, are numerically $0$ as expected. This confirms our proposition that $L \cdot \textbf{1} = \textbf{0}$ holds for both Scenarios 1 and 2.
}

\renewcommand{\arraystretch}{1.3}
\begin{table}[htbp!]
  \centering
  \tabcolsep=0.50 cm
    \caption{Numerical Validation of \textbf{Lemma 1}}
  \label{table:L1}
  \vspace{0.2cm}
\begin{tabular}{cc}
\hline
\multicolumn{2}{c}{\bf{Mean $\pm$ Std. value for elements in $L \cdot \textbf{1}$}   }                                              \\ \hline
\multicolumn{1}{c||}{\textbf{Scenario 1}} & \textbf{Scenario 2}  \\ \hline
\multicolumn{1}{c||}{$(0.67\,\pm\, 2.66)\times 10^{-15}$}           &   $(0.94\, \pm \,2.56)\times 10^{-14}$                 \\ \hline \\
\end{tabular}
\\
\begin{tabular}{cc}
\hline
\multicolumn{2}{c}{\bf{Maximum value for elements in $L \cdot \textbf{1}$}   }                        \\ \hline
\multicolumn{1}{c||}{\textbf{Scenario 1}} & \textbf{Scenario 2} \\ \hline
\multicolumn{1}{c||}{$1.07 \times10^{-14}$ }           &    $9.95 \times10^{-14}$        \\ \hline
\end{tabular}
\end{table}
\section{Conclusions}
\label{sec:conclusion}
This paper focuses on the 
the impact of replacing SGs with GFM-based resources 
on the inter-area oscillation patterns in transmission networks. Theoretical analysis and numerical observations based on small-signal models of inverters and SGs have shown that such impacts can be significant in changing the long-perceived coherency structures of the system. Insights drawn from the study can aid in system planning and can avert the potential risks originating from poorly-damped low-frequency oscillations.
\balance
\bibliography{ref}
\bibliographystyle{IEEEtran}
\end{document}